\documentclass[pra,twocolumn,letterpaper,showpacs,superscriptaddress,floatfix]{revtex4}
\usepackage{graphicx,psfrag,amsmath,amssymb,amsfonts,bbm,latexsym,color,dcolumn,bm}

\begin{document}

\title{Predictability sieve, pointer states, and the classicality of quantum trajectories} 

\author{D.A.R. Dalvit}
\affiliation{Theoretical Division, MS B213, Los Alamos National 
Laboratory, Los Alamos, NM 87545, USA}

\author{J. Dziarmaga}
\affiliation{Institute of Physics and Center for Complex Systems, 
Jagiellonian University,
Reymonta 4, 30-059 Krak\'{o}w, Poland} 

\author{W.H. Zurek}
\affiliation{Theoretical Division, MS B213, Los Alamos National 
Laboratory, Los Alamos, NM 87545, USA}

\date{\today} 

\begin{abstract} 

We study various measures of classicality of the states of open quantum systems subject to 
decoherence.  Classical states are expected
to be stable in spite of decoherence, and are thought to leave conspicuous imprints on the
environment. Here these expected features of environment-induced superselection (einselection)
are quantified using four different criteria: predictability sieve (which selects states that produce
least entropy), purification time (which looks for states that are the easiest to find out from the 
imprint they leave on the environment), efficiency threshold (which finds states that can be
deduced from measurements on a smallest fraction of the environment), and purity loss time 
(that looks for states for which it takes the longest to lose a set fraction of their initial purity). 
We show that when pointer states -- the most predictable states of an open quantum system selected
by the predictability sieve -- are well defined, all four criteria agree that they are indeed
the most classical states.
We illustrate this with two examples: an underdamped harmonic oscillator, for which coherent
states are unanimously chosen by all criteria, and a free particle undergoing quantum Brownian
motion, for which most criteria select almost identical Gaussian states (although, in this case, predictability  sieve does not select well defined pointer states.) 

\end{abstract} 


\maketitle 

\section{Introduction}

Persistent monitoring of an open quantum system by its environment can single out a preferred set
of states, known as pointer states. Pointer states are the most robust in spite of the interaction with the
environment. That it, they entangle least with the environment, and, hence, are
least perturbed by it \cite{zurek81,zurek82}. This is the essence of environment-induced 
superselection or einselection. 

In the standard treatment of environment-induced decoherence, the 
search for pointer states is based on the unconditional master equation for the reduced density matrix 
of the system \cite{zurek93,coherent,gallis}.  
Unconditional dynamics is obtained after tracing over the environment, 
{\it i.e.}, by discarding any information about the system present in the environment. 
Predictability  sieve, introduced in \cite{zurek93,coherent}, explored and extended
in \cite{gallis,tegmark}, and reviewed in \cite{leshouches, joos,rmp},
is a way to quantify predictability and, hence, classicality of states:
For every  pure initial state $|\Psi \rangle$ of the system one asseses the loss of predictability caused by the environment 
by computing entropy $H_{|\Psi\rangle} = - {\rm Tr} \rho_{|\Psi\rangle}(t) \log \rho_{|\Psi\rangle}(t)$, 
or some other measure of predictability (e.g., purity ${\rm Tr}(\rho_{|\Psi\rangle})^2$) from the reduced  
density matrix $\rho_{|\Psi\rangle}(t)$ (which starts as 
$\rho_{|\Psi\rangle}(0) = |\Psi\rangle\langle \Psi|$). Entropy is a function of time and a functional 
of the initial state $|\Psi \rangle$. Pointer states are obtained by minimizing $H_{|\Psi\rangle}$
over $|\Psi\rangle$ and demanding that the answer be robust when varying $t$ within a reasonable
range.

For example, in an underdamped harmonic oscillator coherent states are the pointer states \cite{coherent}. 
This means that for any time longer than a fraction of the period of the oscillator, starting the evolution 
from a coherent state (or some slightly squeezed state that is very close to a coherent state) maximizes 
predictability. For very short times ({\it i.e.},  times short compared to the oscillator period) 
states squeezed in position  tend to do better when the oscillator is 
coupled to the environment through position. These subtleties disappear when the rotating wave approximation 
is adopted, and we shall do so here. But, generally, for such short times (or, alternatively, for a free 
particle) equally unambiguous pointer states do not exist: the most predictable states tend to depend significantly on the time $t$ for which the prediction is needed.

Recently, it has been recognized that one finds out states of systems indirectly by measuring
the environment \cite{zurek93}. This recognition elevates the role of the environment to a communication channel  
and to a witness of the state of the system \cite{rmp}. The dynamics of this process can be analyzed
using unravellings of the unconditional master equation for the density matrix of the system
\cite{wisvac98,wisbra00,ourprl,wisvac02b,ourpra,wiseman}. This is not the only possible approach: information theory can be employed
to analyse the role of the environment as a witness perhaps even more rigorously
\cite{rmp,zurekannalen,olivier1,robin1}. 

The program of quantum unravellings was initiated some time ago and
developed primarily in the context of quantum optics \cite{unravellings}. More recently, it has been applied 
to study the problem of decoherence and the quantum/classical transition  \cite{wisvac98,wisbra00,ourprl,wisvac02b,ourpra,habib,wiseman}.
For a fixed system-environment interaction, 
different quantum unravellings correspond to different measurements on the environment. 

Our work \cite{ourprl,ourpra}  suggested that pointer states should be the easiest to find out from
the environment: The same measurement scheme whose records correlate with the pointer states
is the one that has the shortest purification time of any initial mixed state. This, as well as related
results \cite{olivier1}, lead one to conjecture that different classicality criteria (such as predictability
sieve and purification time) may lead to the same pointer states. 
This conclusion is also of obvious interest in applications (e.g., quantum control) where one would
like to tailor the measurement strategy to optimize some criteria of performance. 

In this paper we will
analyze further different classicality criteria, some of them considered recently in Refs.\cite{wisvac02b,wiseman},
and we will show that when pointer states are well defined, all these criteria are indeed optimized by 
a single strategy for monitoring the environment (that always selects the same pointer states of the
system).  When pointer states are imperfect,  different criteria are optimized by different measurement schemes. 
However, at least in the examples we have  encountered, all these schemes still single out almost identical, most classical states. 

We consider two models. We first study an underdamped harmonic oscillator coupled to a zero temperature 
and to a finite temperature harmonic oscillator bath.  As noted earlier, in this case coherent states are pointer states. We shall show 
that a single unravelling -- heterodyne detection --  optimizes all four classicality criteria that we consider. 
We also study the model of a free particle 
undergoing quantum Brownian motion at finite temperature. A free particle does not have equally
well defined pointer states: they are Gaussians that are approximately position eigenstates at high 
temperature, and less squeezed Gaussians at low temperatures, but the exact ``shape" of the Gaussians depends on $t$. 
We show that, although in this case different classicality criteria are optimized by different 
measurement strategies, the states that are singled out are nearly identical in all cases.

\section{An underdamped harmonic oscillator}

\subsection{Zero temperature bath}

We consider a one dimensional harmonic oscillator with frequency 
$\omega$, position $x$ and momentum $p$ in a zero temperature 
environment of harmonic oscillators linearly coupled to the system 
oscillator. We consider continuous Markovian unravellings with 
homodyne and heterodyne detection schemes for the phonon environment.
In the interaction picture and after the rotating wave approximation,
which assumes that $\omega$ is much larger than the spontaneous
emission rate, the stochastic master equation (SME) describing the 
conditional evolution of the density matrix of the system is
\begin{equation}
d \rho = 
{\cal D}[a] \rho dt  + 
\sqrt{\eta_x} dW_x {\cal H}_{\phi}[a] \rho +
\sqrt{\eta_y} dW_y {\cal H}_{\phi+\frac{\pi}{2}}[a] \rho ,
\label{hetSME}
\end{equation}
where the Lindblad term 
\begin{equation}
{\cal D}[a] \rho = 
a \rho a^{\dagger} - 
\frac{1}{2} a^{\dagger} a \rho - 
\frac{1}{2} \rho a^{\dagger} a ,
\end{equation}
describes damping and decoherence due to spontaneous emission of phonons, 
and
\begin{equation}
{\cal H}_{\phi}[a] \rho = 
a \rho e^{-i \phi} + \rho a^{\dagger} e^{+i \phi} - 
\rho {\rm Tr} [ a \rho e^{-i \phi} + \rho a^{\dagger} e^{+ i \phi} ] ,
\end{equation}
is a nonlinear ``discovery" term that feeds back into the master equation 
information about the state of the system gained from measuring the 
environment. We use units where the spontaneous emission time is $1$.
Here $\eta=\eta_x+\eta_y$ is the measurement efficiency, 
$\phi$ is the phase of the local homodyne oscillator, $dW_x$ and $dW_y$
are uncorrelated Gaussian Wiener increments, $\overline{dW_x}=\overline{dW_y}=0$, 
$\overline{dW^2_x}=\overline{dW^2_y}=dt$, and $\overline{dW_x dW_y}=0$. 
The choice $\eta_x=\eta$ and $\eta_y=0$ corresponds to homodyne unravelling, while 
$\eta_x=\eta_y=\frac12\eta$ to  heterodyne detection \cite{wiseman2,wiseman3}. 
As the harmonic oscillator Hamiltonian $H$ is invariant under rotations in the
$x-p$ plane from now on we set $\phi=0$ without loss of generality.
After fixing $\phi$ the measurement scheme is fully determined by the pair 
of numbers $(\eta_x,\eta_y)$.

It is useful to transform Eq.(\ref{hetSME}) to an equation for the Wigner 
probability distribution ${\cal W}(x,p)$:
\begin{eqnarray}
d {\cal W}(x,p) &=&   
dt \hat{D} {\cal W}+
\sqrt{\eta_x} dW_x [ \hat{\cal H}_{0}  {\cal W} - 
                     {\cal W} {\rm Tr} ( \hat{\cal H}_{0} {\cal W} ) ] + \nonumber \\
&&
\sqrt{\eta_y} dW_y [ \hat{\cal H}_{\frac{\pi}{2}} {\cal W} - 
                     {\cal W} {\rm Tr} ( \hat{\cal H}_{\frac{\pi}{2}} {\cal W} ) ] , 
\end{eqnarray}
where 
\begin{equation}
\hat{\cal D} {\cal W} = 
\left( 
1 + 
\frac{1}{2} (x \partial_x + p \partial_p) + 
\frac{1}{4} (\partial_x^2 + \partial_p^2) 
\right) {\cal W}(x,p) dt ,
\end{equation}
and 
\begin{eqnarray}
\hat{\cal H}_{\phi} {\cal W} &=& 
\sqrt{2} 
\left( 
x \cos \phi + p \sin \phi + \right. \nonumber \\
&& \left. \frac{1}{2} ( \cos \phi \; \partial_x + \sin \phi \; \partial_p) 
\right) {\cal W}(x,p) .
\end{eqnarray}
Since the unconditional ($\eta=0$) steady-state has a Gaussian Wigner distribution, and the
conditional evolution preserves Gaussianity, we propose a solution of the form
\begin{equation}
{\cal W}(x,p) = e^{ -\alpha(t) [x-x_0(t)]^2 - \beta(t) [p-p_0(t)]^2 + \delta(t)   }.
\label{wigner}
\end{equation}
It is also worth noting that there exists a general argument that the most predictable states ought
to be Gaussians \cite{zurek93,coherent}.
The inverse of the variances in position, $\alpha(t)$, and in momentum, $\beta(t)$, evolve 
deterministically even under the conditional stochastic evolution:
\begin{eqnarray}
\alpha(t) &=& 
\frac{e^t [\alpha_0 (1-\eta_x) + \eta_x] - (1-\alpha_0) \eta_x}
     {e^t [\alpha_0 (1-\eta_x) + \eta_x] + (1-\alpha_0) (1-\eta_x)} , \nonumber \\
\beta(t) &=& 
\frac{e^t [\beta_0 (1-\eta_y) + \eta_y] - (1-\beta_0) \eta_y}
     {e^t [\beta_0 (1-\eta_y) + \eta_y] + (1-\beta_0) (1-\eta_y)} ,
\label{abSME}
\end{eqnarray}
where $\alpha_0$ and $\beta_0$ are the initial conditions. In particular, the 
unconditional evolution with $\eta_x=\eta_y=0$, 
\begin{eqnarray}
\alpha(t) &=&  \frac{1}{1 - e^{-t} ( 1- 1/\alpha_0)}  ,\nonumber \\
\beta(t) &=&  \frac{1}{1 - e^{-t} ( 1- 1/\beta_0)}  ,
\end{eqnarray}
relaxes the state to vacuum on a timescale 
$t \simeq {\rm max}\left[\ln(1/\alpha_0),\ln(1/\beta_0)\right]$. For an initial thermal
state with $\alpha_0=\beta_0<<1$ this relaxation time is much larger than $1$. 

\begin{figure}
\vspace{1cm}
\includegraphics[width=8.6cm]{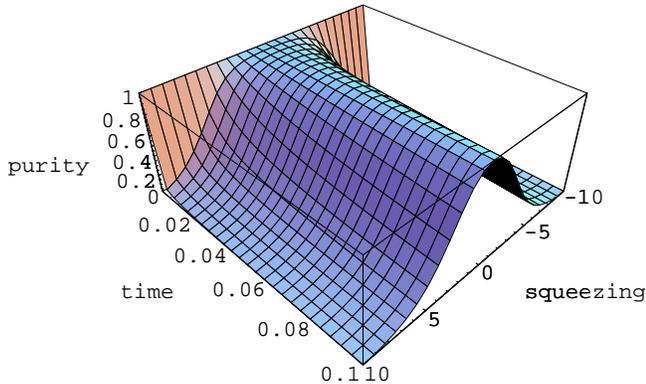}
\caption{Purity as a function of time under unconditional evolution at $T=0$ 
starting from an initial pure Gaussian state parametrized by a squeezing parameter $\xi=\log \kappa$. 
Purity is  not lost for $\xi=0$, {\it i.e.}, the coherent state with $(\alpha,\beta)=(1,1)$. Coherent states are perfect pointer states at $T=0$, and optimize
the predictability sieve criterion. By contrast, purity is initially lost on a decoherence time scale for other states with $\xi \neq 0$. 
Time is measured in units of the spontaneous emission time.}
\label{sieve0}
\end{figure}

The first criterion we consider is predictability sieve.
It is well known that for  an  underdamped harmonic oscillator at zero 
temperature, preferred states - as selected by the predictability sieve - 
are coherent states \cite{zurek93,coherent}: these are the initial pure states that minimize 
the rate of purity loss. In fact, when $T=0$ they are perfect pointer states in the sense that an initial 
coherent state ($\alpha_0=\beta_0=1$) remains a pure coherent state with $\alpha(t)=\beta(t)=1$, see 
Eqs.(\ref{abSME}), and the purity loss is identically zero \cite{goetsch1,goetsch2,giovannetti}. 
In Fig. (\ref{sieve0}) we plot purity $P(t) = \sqrt{\alpha(t) \beta(t)}$ for zero temperature under unconditional evolution starting from initial pure states parametrized by one parameter $\kappa$ as
$\alpha(0)=\kappa$, $\beta(0)=1/\kappa$. As already mentioned, the predictability sieve criterion is optimized for initial coherent states ($\kappa=1$).
Now we will show that these states are also the most classical states according to three different
criteria: purification time, efficiency threshold, and purity loss time. 
We will also show that a single 
measurement strategy,  heterodyne detection, maximizes the robustness of all those 
criteria. This is expected, since  heterodyne detection measures the $x$ and $p$ 
quadratures of the state on an equal footing, and a coherent state has equal variances 
of position and momentum.

The second criterion we consider is purification time. We start from a thermal 
state with $\alpha_0=\beta_0 \ll 1$, and study how fast the state is purified by the 
measurement. Purity $P(t) = \sqrt{\alpha(t) \beta(t) }$ can be shown to evolve 
deterministically, so there is no need to do conditional stochastic simulations to study 
purification time. The initial rate of purity gain can be easily evaluated
\begin{equation}
\left. \frac{dP}{dt} \right|_{t=0} = \sqrt{\eta_x \eta_y} .
\end{equation}
It is maximized for  heterodyne detection $\eta_x=\eta_y=\eta/2$. 
To go beyond this simple calculation of the initial purification rate, we calculate
purification time, defined as the time needed by the conditional
evolution to increase purity to $P=0.5$, which is half way between the initial thermal value $P=0$,
and the final pure state $P=1$ \cite{wiseman}.  In Fig.(1) we plot purification time for different values of the
global efficiency $\eta$ as a function of the measurement scheme,
parametrized by $s\in[0,1]$: $\eta_x=\eta\cos^2\left(s\frac{\pi}{2}\right)$ and
$\eta_y=\eta\sin^2\left(s\frac{\pi}{2}\right)$. As expected, purification time is minimal for 
heterodyne detection ($s=0.5$), which corresponds to a coherent state ({\it i.e.}, a  pointer state). 

\begin{figure}
\vspace{1cm}
\includegraphics[width=8.6cm]{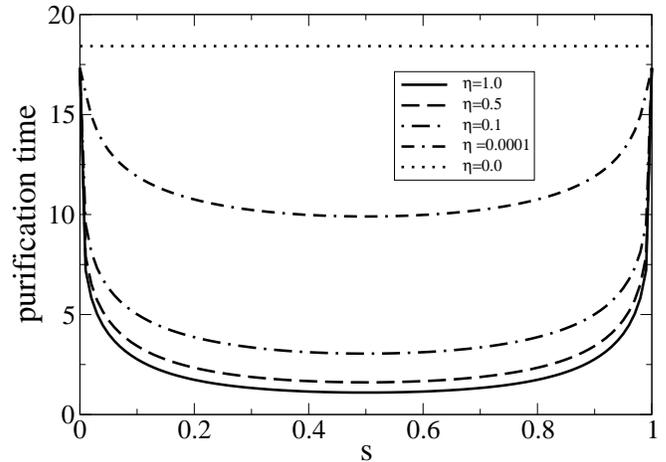}
\caption{Purification time as a function of the measurement scheme parametrized by
$s\in[0,1]$: $\eta_x=\eta\sin^2\left(s\frac{\pi}{2}\right)$ and $\eta_y=\eta\cos^2\left(s\frac{\pi}{2}\right)$.
The initial state is a fat Gaussian state  $\alpha_0=\beta_0=10^{-8}$.  The temperature of the bath
is zero. We note
that while there is a clear optimum in Fig. (1), the very best state $s=0.5$ is not all that much better
than $s \sim 0.2$ or $0.8$. Thus, there is a large ``quantum halo" \cite{halo} of states that maintain purity well
(and in this sense are reasonably classical).
}
\label{fig1}
\end{figure}

\begin{figure}
\vspace{1cm}
\includegraphics[width=8.6cm]{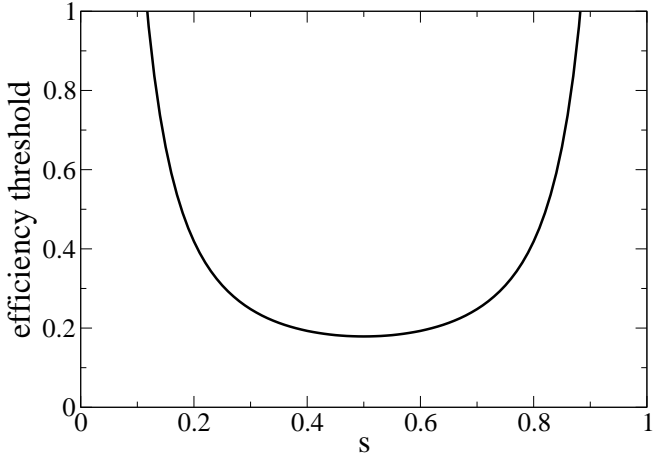}
\caption{ The minimal efficiency threshold $\eta_{\rm thr}$ needed to
reach the threshold purity $P_{\rm thr}=0.5$ after the time
$t_{\rm thr}=2.5$ as a function of the measurement scheme parametrized
by $s\in[0,1]$: $\eta_x=\eta_{\rm thr} \sin^2\left(s\frac{\pi}{2}\right)$ and
$\eta_y=\eta_{\rm thr} \cos^2\left(s\frac{\pi}{2}\right)$. The initial state and temperature are the same as in Fig. 1. Again, $s=0.5$ is surrounded by a sizable ``halo" of (nearly) pointer states \cite{halo}.}
\label{fig2}
\end{figure}

The third criterion is efficiency threshold: the minimal efficiency
$\eta_{\rm thr}$ needed to reach a threshold value of purity $0<P_{\rm
thr}<1$ starting from a mixed high temperature thermal state.  This criterion
cannot be directly used here to discriminate between different
measurements because in a long time the state always relaxes to the
pure vacuum state no matter what the efficiencies $\eta_x$ and
$\eta_y$ are, even in the unconditional evolution
\footnote{This is, incidentally, why coherent states at $T=0$ are not as perfect pointer states
as the eigenstates of the perfect pointer observable $\Lambda$ that commutes with both the self
Hamiltonian of the system, and with the system-environment interaction Hamiltonian
\cite{zurek82}. The coherent
pointer state eventually forgets its initial setting (see Fig. 1), so it could not be a good pointer of the apparatus,
in contrast to eigenstates of a perfect pointer observable.}.
When applied literally, the 
criterion always gives the trivial result $\eta_{\rm thr}=0$. 
This is why we study a modified criterion: we measure the minimal $\eta_{\rm thr}$ 
needed to reach a $P_{\rm thr}$ after a time $t_{\rm thr}$. In Fig. \ref{fig2} we
show $\eta_{\rm thr}$ needed to reach $P_{\rm thr}=0.5$ after
$t_{\rm thr}=2.5$ for different measurement schemes parametrized by $s\in[0,1]$:
$\eta_x=\eta\cos^2\left(s\frac{\pi}{2}\right)$ and
$\eta_y=\eta\sin^2\left(s\frac{\pi}{2}\right)$. As expected, the $\eta_{\rm
thr}$ is minimal for  heterodyne detection ($s=0.5$).

The fourth criterion is the purity loss time (proposed in \cite{wiseman}, and called 
``mixing time" in that reference). The conditional trajectory has to evolve long enough for the state to become pure. Then one stops measuring the environment to see how fast the purity of the state
decreases as a function of time, for different initial measurement schemes. 
All this has to be done before the state relaxes to vacuum. As already mentioned,
for a high temperature initial state with $\alpha_0=\beta_0<<1$ the unconditional evolution
relaxes the state to vacuum after time $\ln(1/\alpha_0) \gg 1$. In the limit of $\alpha_0\to 0$
this relaxation time tends to infinity. In the same limit the conditional evolution 
becomes
\begin{eqnarray}
\alpha(t) &=& 
\frac{e^t-1   }
     {e^t-1+\frac{1}{\eta_x}   } , \\
\beta(t) &=& 
\frac{e^t-1}
     {e^t-1+\frac{1}{\eta_y}   } , 
\end{eqnarray}
and the purity $P(t)=\sqrt{\alpha(t) \beta(t)}$ approaches $1$ after a purification time
${\rm max} \left[\ln(1/\eta_x),\ln(1/\eta_y)\right]$. When both $1 \ge \eta_x, \eta_y \gg \alpha_0$,
then the purification time is much shorter than the relaxation time (compare 
also Fig.\ref{fig1}), and the conditional state collapses to a coherent state much 
faster than the relaxation time in the unconditional evolution. To measure the purity
loss time once the pure coherent state is achieved, the evolution is switched to the 
unconditional evolution with $\eta_x=\eta_y=0$, and the purity decay is observed. 
However, since all measurement strategies collapse the state to a coherent state, the 
subsequent  unconditional evolution does not destroy the purity of this state: the 
amplitude of the pure coherent state smoothly relaxes to the vacuum. This infinite 
purity loss time is the same for all measurement schemes. Coherent states, in the case
of $T=0$,  are perfect 
pointer states: once the conditional state is collapsed to a coherent state it cannot 
lose any purity.

\subsection{Finite temperature bath}

The SME for the harmonic oscillator coupled to a finite temperature bath at temperature $T$ is
\begin{eqnarray}
d \rho &=& (n+1) D[a] \rho dt + n D[a^{\dagger}] \rho dt + \nonumber \\
&&  
\frac{\sqrt{\eta_x}}{\sqrt{1+2n}}  dW_x  \left\{  {\cal H}_{\phi}[(n+1) a] \rho - {\cal H}_{\phi} [ n a^{\dagger}] \rho  \right\} + \nonumber \\
&&
\frac{\sqrt{\eta_y}}{\sqrt{1+2n}}  dW_y  \left\{  {\cal H}_{\phi+ \frac{\pi}{2}}[(n+1) a] \rho - {\cal H}_{\phi- \frac{\pi}{2}} [ n a^{\dagger}] \rho  \right\} \nonumber ,
\end{eqnarray}
where $n=(e^{\hbar \omega_0 / k_{\rm B} T} -1 )^{-1}$ is the Bose distribution. This equation is valid
in the interaction picture and after the rotating wave approximation is performed.
We can write the
analogous equation for the Wigner distribution. We take $\phi=0$, without loss of generality, and get

\begin{eqnarray}
d {\cal W}(x,p) &=&    dt  \hat{D}   {\cal W} + 
\sqrt{\eta_x} dW_x [ \hat{\cal H}_{x}  {\cal W} - 
                     {\cal W} {\rm Tr} ( \hat{\cal H}_{x} {\cal W} ) ] + \nonumber \\
&&
\sqrt{\eta_y} dW_y [ \hat{\cal H}_{y} {\cal W} - 
                     {\cal W} {\rm Tr} ( \hat{\cal H}_{y} {\cal W} ) ] , 
\end{eqnarray}
where now
\begin{equation}
\hat{\cal D} {\cal W} = 
\left( 
1 + 
\frac{x \partial_x + p \partial_p}{2} + 
\frac{1+2 n}{4} (\partial_x^2 + \partial_p^2) 
\right) {\cal W}(x,p) dt ,
\nonumber
\end{equation}
and 
\begin{eqnarray}
\hat{\cal H}_{x} {\cal W} &=& 
\sqrt{2} 
\left( x   + \frac{1}{2} (1+2 n)  \partial_x  \right) {\cal W}(x,p) ,
\nonumber \\
\hat{\cal H}_{y} {\cal W} &=& 
\sqrt{2} 
\left( p  + \frac{1}{2} (1+2 n)  \partial_p  \right) {\cal W}(x,p) .
\nonumber
\end{eqnarray}
Assuming a Gaussian Ansatz for the Wigner function we  get the deterministic equations for
$\alpha(t)$ and $\beta(t)$. They are the same equations as for zero temperature, with the
replacement $\alpha(t) \rightarrow (1+2n) \alpha(t)$. Hence, the solution can be straightforwardly
obtained from Eq.(\ref{abSME}).

\begin{figure}
\vspace{1cm}
\includegraphics[width=8.6cm]{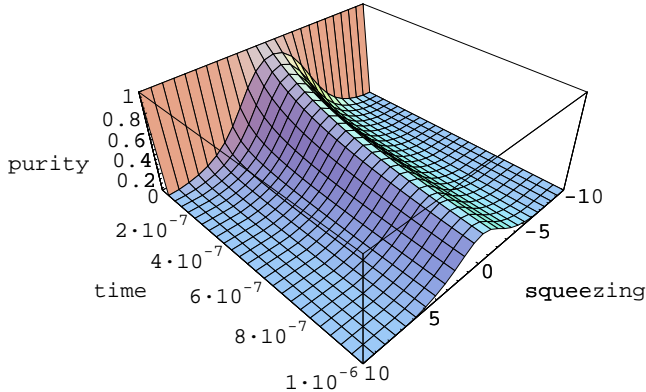}
\caption{Purity as a function of time under unconditional evolution at finite temperature
starting from an initial pure Gaussian state parametrized by the squeezing parameter 
$\xi=\log \kappa$.  The temperature of the bath is $k_{\rm B}T / \hbar \omega_0=10^6$,
where $\omega_0$ is the frequency of the system oscillator. 
Again, coherent states ($\xi=0$) are selected by predictability sieve, as they lose purity most
slowly. For large times, all initial states lead to the same stationary state with 
purity $P(\infty) = 1/(1+2n)$, where $n \approx 10^6$ for this temperature. Time is measured
in units of the spontaneous emission time.
}
\label{sieve6}
\end{figure}

When $T>0$, the stationary state of the evolution is  $\alpha_{\rm ss} = \beta_{\rm ss} = 1/(1+2n)$, 
{\it i.e.}, a Gaussian with variances in position and momentum larger than 1. The unconditional
evolution relaxes to this stationary state on a time scale $t \simeq {\rm max} [ \ln(1/(1+2n) \alpha_0) ,
\ln( 1/(1+2n) \beta_0 )]$. An initial pure coherent state $\alpha_0=\beta_0=1$ does not remain pure;
it loses purity and reaches the stationary state with final purity $P(t\rightarrow \infty) = 1/ (1+2n)< 1$. 
Predictability sieve still selects coherent states as pointer states, even at finite temperature
(see Fig. \ref{sieve6}). 
Also,   heterodyne detection measurement  maximizes the robustness of all the criteria
(purification time, efficiency threshold, and purity loss time).  For an initial high temperature
state such that $(1+2n) \alpha_0 = (1+2n) \beta_0  \ll1$, the unconditional evolution relaxes  
towards the stationary state on a very large time scale with $P(\infty)=1/(1+2n)$, while  the 
conditional evolution reaches the same asymptotic value of purity on a much shorter time scale, 
of the order of ${\rm max} [ \ln (1/\eta_x), \ln(1/\eta_y) ]$.  At finite temperature,  the purity loss time  is not infinity as for $T=0$: all measurement schemes lead to the same, non pure state, then the measurement is switched off, and then purity is reduced again to the same final value, irrespective of the measurement scheme. This criterion cannot distinguish measurement 
schemes.

In summary, all classicality robustness criteria select the same states (coherent states) for the model
of an underdamped harmonic oscillator, for all temperatures of the bath.

\section{A free particle undergoing quantum Brownian motion at high temperature}

The high temperature conditional homodyne and heterodyne master equation for 
a free particle undergoing quantum Brownian motion is \cite{diosi,wiseman}
\begin{eqnarray}
d\rho &=& -\frac{i}{2} \left[ p^2+\frac{qp+pq}{2},\rho\right] dt +
{\cal D}[c]\rho dt + \nonumber \\
&& \left[ \sqrt{\eta} \; dW \left( c \rho- \rho {\rm Tr}(\rho c) \right)+
   {\rm h.c.}\right]~. 
\label{SMEQBM}
\end{eqnarray}
Here $\eta$ is the total efficiency of the measurement, and the annihilation operator 
$c$ is
\begin{equation}
c=\frac{1}{\sqrt{2}}\left(\sqrt{4T}q+\frac{ip}{\sqrt{4T}}\right)~.
\end{equation}
We use rescaled units such that the damping rate, the particle mass, Boltzmann constant, and
$\hbar$ are all unity.
The complex, Gaussian Wigner increment has correlators $\overline{dW dW^*}=dt$ and
$\overline{dW^2}=dt \; r e^{i \varphi}$ with $-1 \le r \leq 1$.  It can be written 
in terms of two real, uncorrelated, Gaussian noises $dW_x$ and $dW_y$ as
\begin{equation}
dW = 
e^{+i \varphi /2} \sqrt{\frac{1+r}{2}} \; dW_x + 
e^{-i \varphi/2} \sqrt{\frac{1-r}{2}} \; dW_y ,
\end{equation}
where $\overline{dW_x^2}=\overline{dW_y^2}=dt$, and $\overline{dW_x dW_y} = 0$. In 
this way, the stochastic  part of the master equation Eq.(\ref{SMEQBM}) can be 
expressed as
\begin{equation}
d \rho^{\rm stoch} = 
\sqrt{\eta_x} dW_x \; {\cal H}_{\phi}[c] \rho + 
\sqrt{\eta_y} dW_y \; {\cal H}_{\phi+\frac{\pi}{2}}[c] \rho .
\end{equation}
Here we have defined $\phi = - \varphi/2$, $\eta_x = \eta (1+r)/2$, and 
$\eta_y = \eta (1-r)/2$.  Heterodyne detection corresponds to $r=0$, 
while $r=1$ corresponds to homodyne measurement of the linear combination 
$c e^{i \phi} + c^{\dagger} e^{-i \phi}$ of $q$ and $p$. Hence, $\phi=0$
corresponds to a measurement of $q$, and $\phi=\pi/2$ to a measurement of $p$.
The unconditional version of the master equation (\ref{SMEQBM}) was derived in 
Ref.\cite{diosi} as Markovian approximation to the exact quantum Brownian motion
master equation at high temperature. The equation is valid only when $T\gg 1$.    

As discussed in the Introduction, the free particle undergoing quantum Brownian motion does not
have well defined pointer states for any temperature of the environment. For example, at high $T$
the Lindblad term in the master equation (\ref{SMEQBM})  formally dominates over the free Hamiltonian
part, and predictability sieve would select eigenstates of the annihilation operator $c$, that correspond
to Gaussian states squeezed in position. However, this answer is not robust under variations of the time
$t$ during which the criterion is applied: states squeezed in position have a large dispersion in momentum, 
and the free dynamical evolution will drive the state away from the squeezed eigenstates of $c$.
We will show below that, because of these facts, different classicality criteria are maximized for different 
unravellings. However, even under these circumstances the selected states are very similar Gaussians. 

Just as in the model of the underdamped harmonic oscillator, the conditional dynamics 
Eq.(\ref{SMEQBM}) preserves Gaussianity. We assume a Gaussian ansatz for the Wigner 
function
\begin{eqnarray}
{\cal W}(x,p) &=&  
\exp \left\{ -\alpha(t) [x-x_0(t)]^2 - \beta(t) [p-p_0(t)]^2 - \right. \nonumber \\
&&          \left.   2 \gamma(t) [x-x_0(t)][p-p_0(t)] + \delta(t) \right\} . 
\label{wigner2}
\end{eqnarray}
The coefficients $\alpha(t)$, $\beta(t)$, and $\gamma(t)$ evolve deterministically 
according to the following equations
\begin{eqnarray}
\frac{d \alpha}{dt} &=&
- \frac{\alpha^2}{4T}-4T \gamma^2 +  4T\eta_x A_1^2 + 4T\eta_y A_2^2 , 
\label{dabcdt} \\
 \frac{d \beta}{dt} &=& - 4T \beta^2-\frac{\gamma^2}{4T}+2 \beta-2 \gamma +
\frac{\eta_x}{4T} B_1^2 + \frac{\eta_y}{4T} B_2^2 , \nonumber \\ 
\frac{d \gamma}{dt} &=& - \frac{\alpha \gamma}{4 T}  - 4T \beta \gamma + \gamma - \alpha + 
\eta_x A_1 B_1- \eta_y  A_2  B_2 , \nonumber  
\end{eqnarray}
where
\begin{eqnarray}
A_1 &=&   \gamma \sin\phi-\left(1-\frac{\alpha}{4T}\right)\cos\phi ,\nonumber \\
A_2 &=&    \gamma \cos\phi+\left(1-\frac{\alpha}{4T}\right)\sin\phi , \nonumber \\
B_1 &=& \gamma \cos\phi-\left(1-4T \beta \right)\sin\phi ,\nonumber \\
B_2 &=&  \gamma \sin\phi+\left(1-4T \beta \right)\cos\phi .\nonumber
\end{eqnarray}

\begin{figure}
\vspace{1cm}
\includegraphics[width=8.6cm]{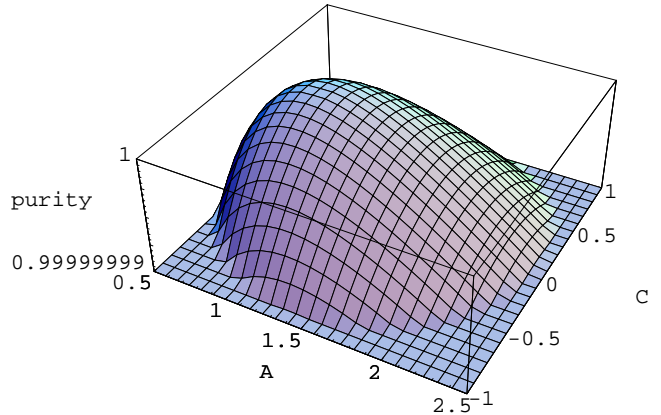}
\caption{ Purity under unconditional evolution at $T=10^6$ after a short time $t=0.1/4T=2.5~10^{-8}$  starting from an initial pure Gaussian state parametrized by two numbers $(A,C)$. 
The parameter $A$ is related to the squeezing  of the probability distribution of the initial state, and
$C$ is related to its tilt in phase space. Purity is 
maximized for an initial state that is approximately  parametrized by $(A,C)=(1,0)$, that is, 
the  eigenstate of the squeezed annihilation operator $c$ with
$(\alpha,\beta,\gamma)=(4T,\frac{1}{4T},0)$, which is the most predictable state after this short 
time of unconditional evolution. However, the vertical range of the plot is only $(0.99999999,1)$ - 
at this early time the unconditional evolution is still very predictable no matter what its
initial pure state is. The most predictable eigenstates of $c$ are not well distinguished from the rest.
Rescaled units are used.
}
\label{PSshort}
\end{figure}

To apply the predictability sieve criterion we solve the unconditional version of these equations
($\eta_x=\eta_y=0$) starting from different initial pure Gaussian states. The initial pure state
is parametrized by two parameters $(A,C)$ as $\alpha(0)=4T~A,\gamma(0)=C$ and 
$\beta(0)=[1+\gamma^2(0)]/\alpha(0)$. In this parametrization the squeezed eigenstate of $c$ is
$(A=1,C=0)$. In figure \ref{PSshort} we show purity $P(t)=\sqrt{\alpha(t) \beta(t) - \gamma^2(t)}$ after unconditional evolution for
a very short time $t=0.1/ 4T=2.5~10^{-8}$ at temperature $T=10^6$ as a function of initial 
pure state $(A,C)$.  Purity is maximal when $(A,C)$ are approximately equal to $(1,0)$, {\it i.e.}, 
purity is maximized for the eigenstate of the squeezed 
operator $c$ with
$(\alpha,\beta,\gamma)=(4T, 1/ 4T, 0)$. As expected, at early times the evolution is dominated 
by the Lindblad term and it is  most predictable for the eigenstates of $c$. In the
high-$T$ limit these eigenstates tend to be  position eigenstates. However, the range of values of purity in figure \ref{PSshort} is only $(0.99999999,1)$, so that after this short time the evolution is still very predictable no matter what the initial state is - the most robust states are not well distinguished from the rest.

\begin{figure}
\vspace{1cm}
\includegraphics[width=8.6cm]{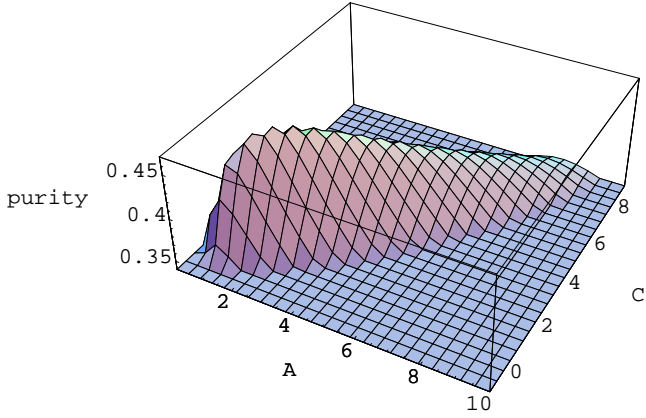}
\caption{ Purity after unconditional evolution at $T=10^6$ for a time $t=2/\sqrt{4T}=10^{-3}$ 
when the unconditional evolution starting from the most robust initial state loses
approximately one half of its initial purity. Purity is maximal when $(A,C)=(1.75,1.75)$, or
$(\alpha,\beta,\gamma)=(1.75\sqrt{4T},\frac{2.32}{\sqrt{4T}},1.75)$. At this time the most 
predictable initial states are well distinguished from (are much more pure than) other states, which already have negligible purity.  Rescaled units are used.
}
\label{PShalf}
\end{figure}

Rather than the early times $t\simeq 1/ 4T$ we found the times $t\simeq 1/ \sqrt{4T}$
more relevant from the point of view of predictability sieve, since then  the unconditional evolution
starting from the most robust initial states loses approximately one half of its initial purity,
while at the same time the evolution starting from bad states is essentially not predictable at all.
At $t\simeq 1/\sqrt{4T}$ the most robust states are well distinguished from ``the Hilbert space chaff" by 
the predictability sieve. 
In figure \ref{PShalf} we show purity $P(t)$ after unconditional evolution at temperature
$T=10^6$ at a time $t=2 / \sqrt{4T} = 10^{-3}$. Purity is maximal when $(A,C)=(1.75,1.75)$ or
$(\alpha,\beta,\gamma)=(1.75\sqrt{4T},\frac{2.32}{\sqrt{4T}},1.75)$. In the high $T$ limit
these states are very different from the eigenstates of $c$ although they also tend to
approximate position eigenstates for large values of the temperature.
In a strict sense predictability sieve does not select any well 
defined and $t$-independent pointer states. Nevertheless, non-local superpositions are still
quickly destroyed, and lose purity much faster than the stable and more or less localized Gaussians we have described.
Therefore, for many practical purposes the states which are selected
at measurement times $t\simeq 1/\sqrt{4T}$ may play the role of the robust classical
states. Parameters of these states scale with $T$ as $\alpha\simeq \sqrt{4T}$, 
$\beta\simeq\frac{1}{\sqrt{4T}}$, and $\gamma\simeq 1$. In the following we will see
that this class of states is also preferred by the other classicality criteria. 

\begin{figure}
\vspace{1cm}
\includegraphics[width=8.6cm]{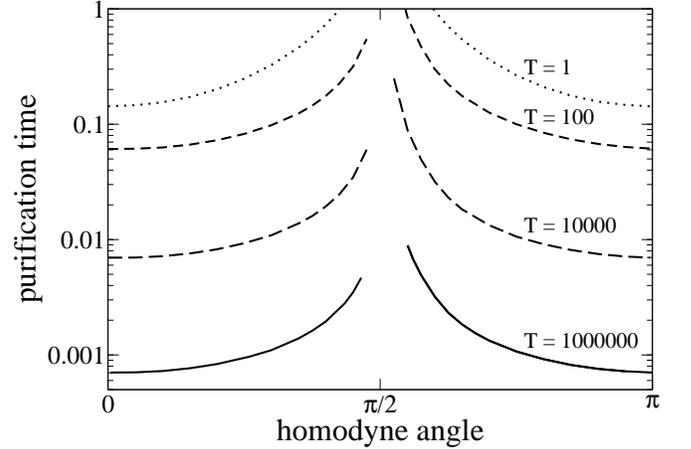}
\caption{ Purification time as a function of the homodyne angle $\phi$ ($r=1$) for
an initial thermal state with temperature $T$. Purification time is the shortest for
$x-$homodyne, {\it i.e.}, when $\phi=0,\pi$. By contrast, $y-$homodyne
($\phi=\pi/2$) does not purify the state at all. Rescaled units are used. }
\label{purificationQBM}
\end{figure}

Now we turn to the second classicality criterion, purification time.
In order to quantify it we prepare the system in its
unconditional stationary thermal state, $\alpha_0=\gamma_0=0, \beta_0=1/2T$. Then
we compute how fast the conditional state is gaining purity $P(t)$  depending on the measurement scheme on the environment, defined by $r$ and $\phi$. Here we assume full efficiency ($\eta=1$). The initial rate of gain of purity can be easily calculated analytically at $t=0$:
\begin{eqnarray}
2 P(0) \dot{P}(0) =
\left.
\left( \beta \frac{d \alpha}{dt} + \alpha \frac{d\beta}{dt} - 
       2 \gamma \frac{d\gamma}{dt} 
\right)
\right|_{t=0} = 
1+r\cos2\phi 
\nonumber
\end{eqnarray}
and it is found not to depend on $T$. This initial purification rate is  fastest
when $r=1$ and $\phi=0,\pi$, or when $r=-1$ and $\phi=\pi/2, 3\pi/2$. All these four
solutions correspond, in fact, to the same $x-$homodyne. The slowest initial
purification is obtained for $y-$homodyne with $r=1,\phi=\pi/2$. 
To go beyond the initial purification we measure the purification time, defined as
the time needed by the fully efficient ($\eta=1$) conditional evolution to increase
purity to $P=0.5$, which is half way between the initial thermal state value ($P \approx 0$)
and the final pure state value ($P=1$) \cite{wiseman}. Purification times for different homodyne
angles $\phi$ at different temperatures are shown in figure \ref{purificationQBM}.  
As expected, purification time is the shortest for $x-$homodyne, {\it i.e.}, when
$r=1$ and $\phi=0,\pi$. In contrast, $y-$homodyne does not purify the state at all.

\begin{figure}
\vspace{1cm}
\includegraphics[width=8.6cm]{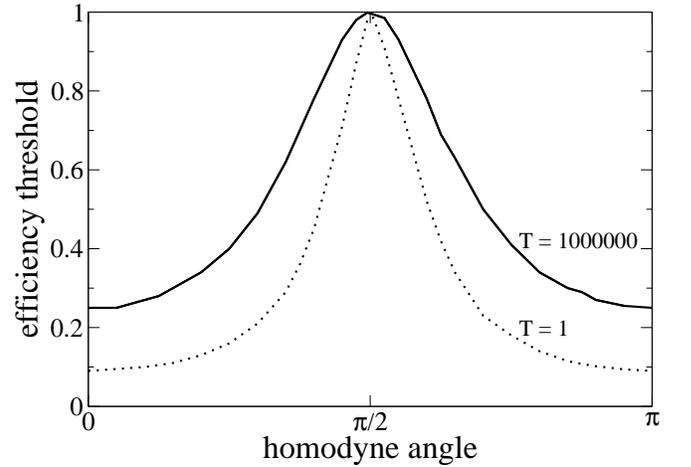}
\caption{ Efficiency threshold as a function of the homodyne angle $\phi$
for an initial thermal state with temperature $T$. The
threshold is the lowest for $x-$homodyne ($\phi=0,\pi$) and the highest for
$y-$homodyne ($\phi=\pi/2$). Rescaled units are used.}
\label{thresholdQBM}
\end{figure}

Now we consider the efficiency threshold, defined as the minimal efficiency
$\eta_{\rm thr}$ required for the conditional evolution to asymptotically reach 
a threshold purity $P_{\rm thr}$. Here we assume $P_{\rm thr}=0.5$ and an initial
thermal state. Efficiency thresholds for different homodyne angles $\phi$ are 
collected in Figure \ref{thresholdQBM}. The threshold is the lowest for $x-$homodyne
($\phi=0,\pi$), and the highest for $y-$homodyne when full efficiency is required
to find out the state after infinitely long time.

Finally, in order to measure the purity loss time (called ``mixing time" in \cite{wiseman}), we evolve the conditional state
of the system for a sufficiently long time so that it reaches a conditional
stationary state, that depends on the measurement scheme. The stationary conditional
state can be obtained by setting the left hand sides of Eqs. (\ref{dabcdt}) to zero
and solving the equations for $\alpha_{\rm ss},\beta_{\rm
ss},\gamma_{\rm ss}$. In the high $T$ limit, the stationary solutions can be found
in the form $\alpha_{\rm ss}= A_{\rm ss} \sqrt{4T}, \beta_{\rm ss}=B_{\rm ss} /
\sqrt{4T}, \gamma_{\rm ss}=C_{\rm ss}$. After keeping in each equation only the
leading order terms in the high $T$ limit we obtain simplified equations for $A_{\rm
ss}, B_{\rm ss}, C_{\rm ss}$, and their solution gives,
\begin{eqnarray}
A_{\rm ss} &=& - \frac{B_{\rm ss} C_{\rm ss}}{2} (1+ r \cos 2\phi) - \frac{B_{\rm ss}}{2} r \sin 2 \phi , \nonumber \\
B_{\rm ss} &=& \left( - \frac{4 C_{\rm ss}}{1 + r \cos 2\phi} \right)^{1/2} , 
\nonumber \\
C_{\rm ss} &=& - \frac{ r \sin 2\phi+ \sqrt{1+2r\cos2\phi+r^2}}{1+r\cos2\phi} .
\end{eqnarray} 
Once the stationary state is achieved the evolution is switched to unconditional
evolution with $\eta=0$ and we calculate the purity loss rate. This depends
on the measurement scheme through the stationary conditional state, which is the
initial state for the unconditional evolution. The initial purity loss rate can be easily 
calculated as
\begin{equation}
 \left. -\frac{dP}{dt} \right|_{t=0} = \sqrt{T} \; B_{\rm ss}, 
\end{equation}
to leading order in high $T$.
This initial  purity loss rate is minimized for $x-$homodyne with $r=1$ and 
homodyne angle $\phi= 5 \pi /6$. 

\begin{figure}
\vspace{1cm}
\includegraphics[width=8.6cm]{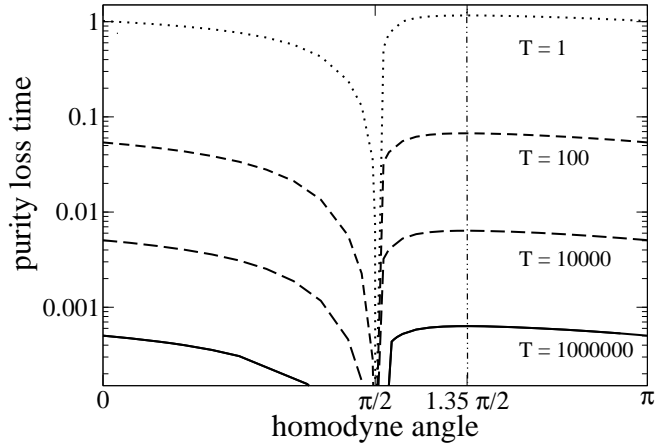}
\caption{Purity loss time needed by unconditional evolution to reduce purity from the
initial value $P=1$ to $P=0.5$, which is half way to the stationary value $P=0$ in
the thermal state. The homodyne angle that maximizes the purity loss time is
$\phi=1.35~\frac{\pi}{2}$. On the other hand, the purity loss time is zero for
$y-$homodyne with $\phi=\pi/2$. Rescaled units are used.}
\label{mixingQBM}
\end{figure}

So far we have analyzed only the initial purity loss time. Now we consider the purity loss time, defined as the time
needed by the unconditional evolution to reduce purity to $P=0.5$, which is half way
between the initial pure state and the stationary thermal state with $P \approx 0$. In
figure \ref{mixingQBM} we show the purity loss  time as a function of homodyne angle
$\phi$ after homodyne measurement with $r=1$. The homodyne angle that maximizes
the purity loss time is $\phi=1.35 \pi / 2$.

\begin{figure}
\includegraphics[width=8cm]{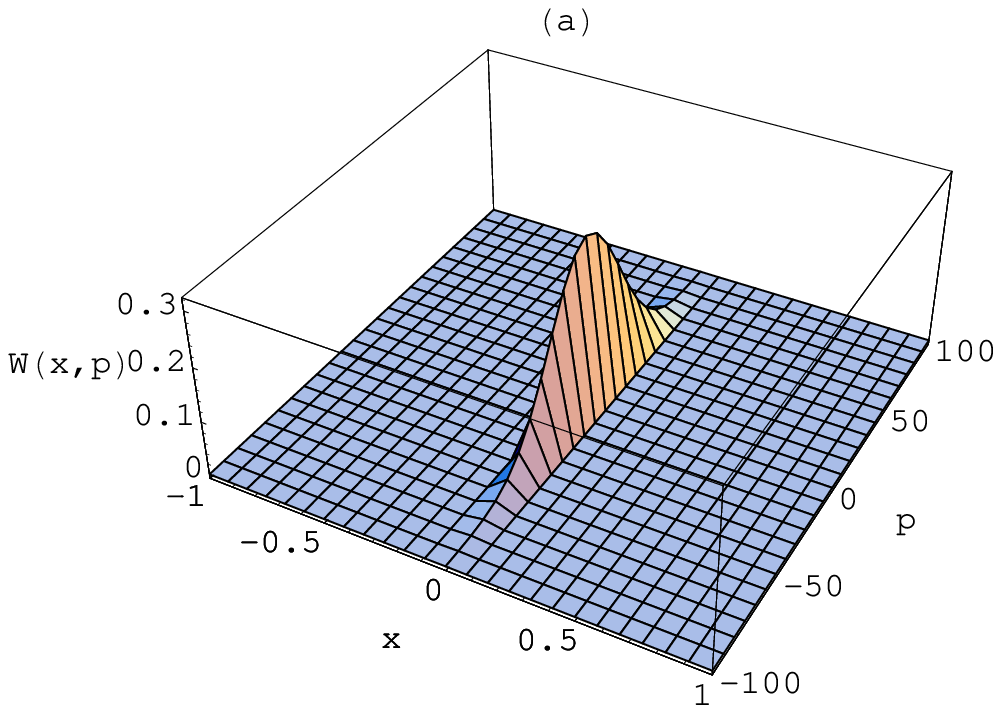}
\includegraphics[width=8cm]{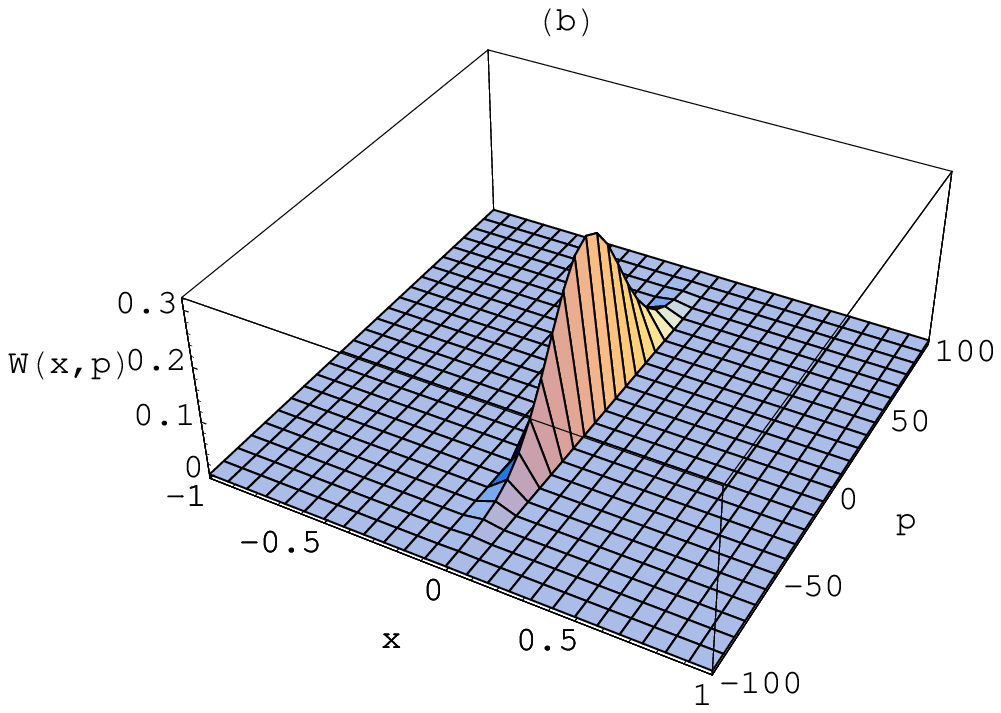}
\caption{Wigner functions of the stationary conditional states selected at
high temperatures $T=10^6$ by different robustness criteria: (a)
purification time and efficiency threshold (homodyne angle $\phi=0$), and
(b) purity loss time (homodyne angle $\phi=1.35 \pi/2$). The two selected
states are very similar (scalar product equal to $0.97$), and
Gaussians squeezed to ``approximate"  position eigenstates. Rescaled units are used.}
\end{figure}

\begin{figure}
\includegraphics[width=8cm]{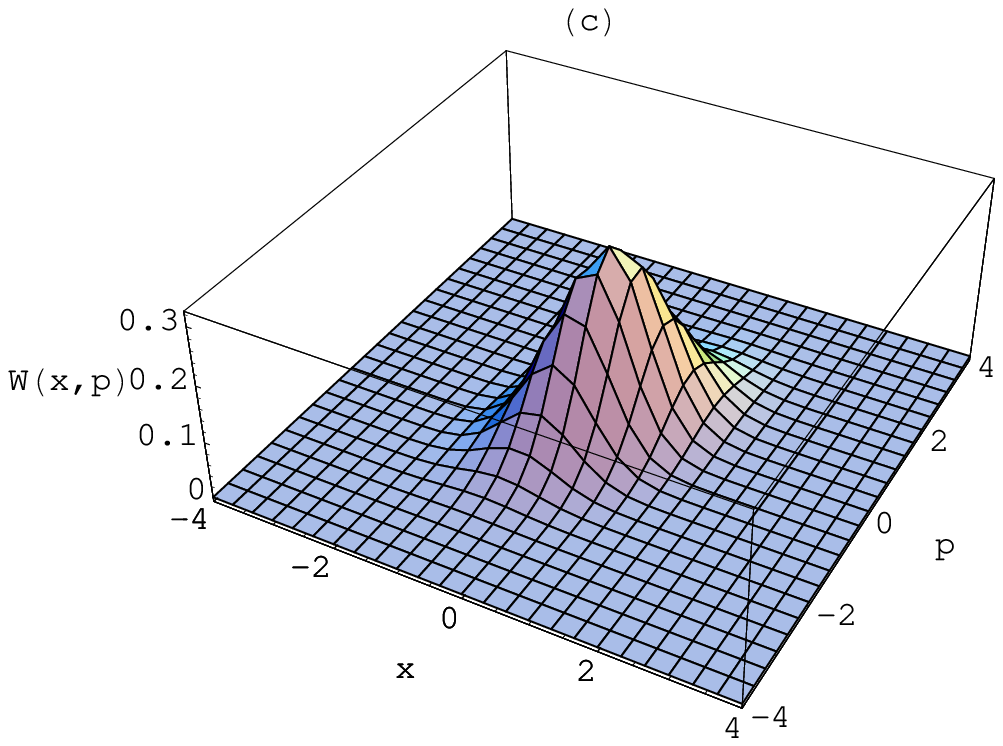}
\includegraphics[width=8cm]{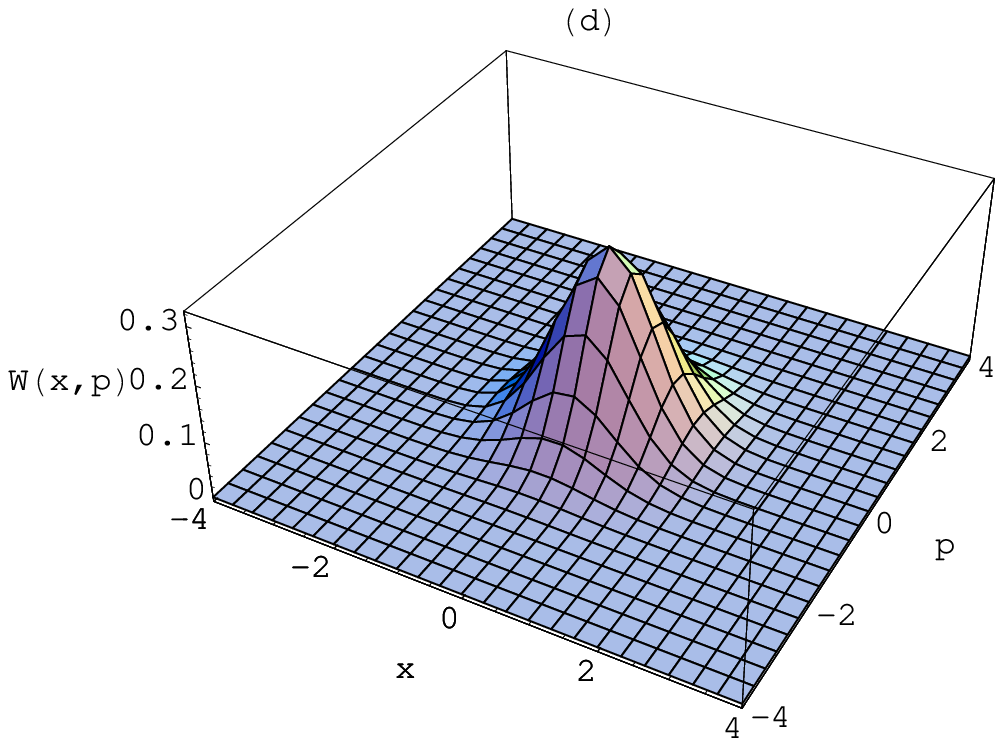}
\caption{Wigner functions of the stationary conditional states selected at
low temperatures $T=1$ by different robustness criteria: (c) purification
time and efficiency threshold (homodyne angle $\phi=0$), and (d) purity loss
time (homodyne angle $\phi=1.35 \pi/2$). The two selected states are very
similar (scalar product equal to $0.99$), and approximately equal to
coherent states with $\alpha=\beta=1$, $\gamma=0$ (scalar product with a coherent state is equal to $0.99$ and 
$0.98$ respectively). Rescaled units are used.} 
\end{figure}

We applied the three different criteria and we found that in two cases -
purification time and efficiency threshold - the most robust measurement
strategy is $x-$homodyne ($\phi=0$) over the whole range of temperatures, but in one
case - purity loss time - the most robust measurement is homodyne with
$\phi=1.35 \pi/2$.   In each case the most robust measurement defines the
most classical state as the conditional stationary state obtained with the
optimal measurement scheme.These results may leave one under the impression that
various criteria lead to different candidates for classical states. Indeed, this was the
conclusion of \cite{wiseman}, where similar results concerning optimal measurement
strategies were obtained for the model of this Section. The key point is, however, that
although different classicality criteria are maximized by different measurement
strategies, in all cases the states that are singled out are almost
identical. 

In the two following tables  we list parameters
of the conditional stationary states for $\phi=0$ and $\phi=1.35 \pi / 2$
for several values of the temperature. The states can be described by
three independent quantities:  the dispersions in position $\Delta x$ and
in momentum $\Delta p$, and the covariance $C_{xp} \equiv \langle x p + p
x \rangle/2 - \langle x \rangle \; \langle p \rangle$. We find that for
all listed temperatures the stationary conditional states for $\phi=0$ and
$\phi=1.35 \pi/2$ are very similar (see figures 6 and 7). 
One can easily imagine they are well within each others ``quantum halo" \cite{halo}.
These states evolve from almost coherent at $T=1$ to
almost position eigenstates at high $T$. Thus, in contrast  to \cite{wiseman}, we conclude
that all of the above criteria lead to essentially the same ``most classical" states.

\begin{table}[htbp]
\centering
\caption{Stationary conditional state for homodyne angle $\phi=0$ at different temperatures $T$. 
Rescaled units are used.}
\label{tabphi0}
\begin{tabular}{|c|c|c|c|c|c|c|}
\hline
 $T$   &   $\alpha_{\rm ss}$  &    $\beta_{\rm ss}$   &  $\gamma_{\rm ss}$   & $\Delta x$ & $\Delta p$ & $C_{xp}$ \\
\hline
$10^6$ & $2826$ & $0.0007$ & $-0.999$ & $0.018$ & $37.968$ & $0.509$ \\
$10^4$ & $280 $ & $0.0070$ & $-0.992$ & $0.059$ & $11.977$ & $0.508$ \\
$10^2$ & $26.4$ & $0.0707$ & $-0.931$ & $0.188$ & $3.633$ & $0.466$ \\
$1   $ & $1.53$ & $0.8002$ & $-0.480$ & $0.632$ & $0.877$ & $0.241$ \\
\hline
\end{tabular}
\end{table}

\begin{table}[htbp]
\centering
\caption{Stationary conditional state for homodyne angle $\phi=1.35 \pi/2$ at different temperatures $T$.
Rescaled units are used. }
\label{tabphi135}
\begin{tabular}{|c|c|c|c|c|c|c|}
\hline
 $T$   &   $\alpha_{\rm ss}$  &    $\beta_{\rm ss}$   &  $\gamma_{\rm ss}$   & $\Delta x$ & $\Delta p$ 
 & $C_{xp}$ \\
\hline
$10^6$ & $1500$ & $0.0007$ & $-0.281$ & $0.018$ & $27.792$ & $0.144$ \\
$10^4$ & $149 $ & $0.0072$ & $-0.280$ & $0.060$ & $8.656$ & $0.141$ \\
$10^2$ & $14  $ & $0.0741$ & $-0.270$ & $0.192$ & $2.694$ & $0.140$ \\
$1   $ & $1.05$ & $0.9561$ & $-0.112$ & $0.694$ & $0.728$ & $0.056$ \\
\hline
\end{tabular}
\end{table}

\section{Conclusions}

In the example of the under-damped harmonic oscillator, where pointer 
states defined by the predictability sieve criterion are unambiguous, we have shown that
different classicality criteria single out  the same states, {\it i.e.}, coherent states that coincide with the pointer states selected by the predictability
sieve \cite{zurek93,coherent}.
We considered three of these  ``alternative classicality criteria": purification time, which looks for states of the system that are the easiest to find out from the imprint they leave on the environment; efficiency threshold, which finds states of the system  that can be deduced from measurements on a smallest fraction of the environment; and purity loss time, that looks for states of the system for which it takes the longest
to lose a set fraction of their initial purity. Our findings indicate that {\it quantum unravelings are in effect classical for pointer states.} 

When pointer states are not well defined, as in the model
of a free particle undergoing quantum Brownian motion, or when pointer states do not exist at all,
different criteria  may select different states. However,  it appears that these candidate
classical states are very similar, defining approximate pointer states, as in the model of the free particle
undergoing quantum Brownian motion. This is an interesting example because different classicality criteria select measurement schemes that appear quite different, but these different schemes turn out to prepare very similar Gaussian states.
 
Following our work \cite{ourprl}, the classical robustness of different quantum unravellings was studied 
in \cite{wiseman}, where some of the classicality measures used in this paper 
-- such as efficiency threshold and purity loss time (called ``mixing time" in \cite{wiseman}) -- were considered.
It was claimed there that for a fixed environmental interaction the level of robustness depends on the
measurement strategy, and that no single strategy is maximally robust in all ways. This conclusion
was drawn from two models: resonance fluorescence of a two-level atom (a model for
which pointer states do not exist), and a free particle in quantum Brownian motion (one of the models used in this paper, for which one might perhaps argue there are approximate, time-dependent, and imperfect pointer states). 
As we discussed above, the conclusion of Ref.\cite{wiseman} does not apply to the case when einselection works well, 
{\it i.e.}, when pointer states are well defined (e.g., the under-damped harmonic oscillator). Moreover, the conclusion of \cite{wiseman}
is misleading  even for the example of the free particle undergoing quantum Brownian motion they have investigated: 
Pointer states are not well defined there, and different robustness criteria
are optimized by different measurement strategies. Yet, the {\it states} resulting from all of the above strategies are essentially identical! Thus, even when
einselection does not pick out  unique pointer states, different criteria still agree on what appears to be most classical. 

Our conclusions may assist in the choice of the optimal measurement strategies, especially in applications that involve quantum control. 
The fact that when good or even approximate pointer states exist, they tend to be prepared by all the seemingly different schemes optimized
for several reasonable criteria, attest to the practical implications of their classicality. 


\section{Acknowledgments}

We are grateful to Howard M. Wiseman for fruitful discussions and correspondence. This work was supported in part by NSA.



\begin{thebibliography}{99} 

\bibitem{zurek81} W.H. Zurek, Phys. Rev. D {\bf 24}, 1516 (1981).

\bibitem{zurek82} W.H. Zurek, Phys. Rev. D {\bf 26}, 1862 (1982).

\bibitem{zurek93} W.H. Zurek,  Prog. Theor. Phys. {\bf 89}, 281 (1993).

\bibitem{coherent} W.H. Zurek, S. Habib, and J.P. Paz, Phys. Rev. Lett. {\bf 70}, 1187 (1993).

\bibitem{gallis} M.R. Gallis, Phys. Rev. A {\bf 53}, 655 (1996).

\bibitem{tegmark} M. Tegmark and H.S. Shapiro, Phys. Rev. E {\bf 50}, 2538 (1994).

\bibitem{leshouches} J.P. Paz and W.H. Zurek, in {\it Coherent Matter Waves}, Les Houches
Summer School, Session LXXII, edited by R. Kaiser, C. Westbrook, and F. David (Springer-Verlag, Berlin, 2001). pp. 533-614.

\bibitem{joos} E. Joos, H.D. Zeh, C. Kiefer, D. Giulini, J. Kupsch, and I.-O. Stamatescu,
{\it Decoherence and the Appearance of a Classical World in Quantum Theory} (Springer, Berlin, 2003).

\bibitem{rmp} W.H. Zurek, Rev. Mod. Phys. {\bf 75}, 715 (2003).

\bibitem{wisvac98} H.M. Wiseman and J.A. Vaccaro, Phys. Lett. A {\bf 250}, 241 (1998).

\bibitem{wisbra00} H.M. Wiseman and Z. Brady, Phys. Rev. A {\bf 62}, 023805 (2000).

\bibitem{ourprl} D.A.R. Dalvit, J. Dziarmaga, and W.H. Zurek, Phys. Rev. Lett. {\bf 86}, 373 (2001).

\bibitem{wisvac02b}  W.H. Wiseman and J.A. Vaccaro, Phys. Rev. A {\bf 65}, 043606 (2002).

\bibitem{ourpra} J. Dziarmaga, D.A.R. Dalvit, and W.H. Zurek, Phys. Rev. A {\bf 69}, 022109 (2004).

\bibitem{wiseman} D.J. Atkins, Z. Brady, K. Jacobs, and H.M. Wiseman, Europhys. Lett. {\bf 69},
 163 (2005).
 
\bibitem{zurekannalen} W.H. Zurek, Annalen der Physik {\bf 9}, 855 (2000).
 
\bibitem{olivier1} H. Ollivier, D. Poulin, and W.H. Zurek, Phys. Rev. Lett. {\bf 93}, 220401 (2004);
also quant-ph/0408125.

\bibitem{robin1} R. Blume-Kohout and W.H. Zurek, quant-ph/0408147; quant-ph/0505031.

\bibitem{unravellings} 
L. Di\'{o}si, Phys. Lett. A {\bf 129}, 419 (1988);
A. Barchielli and V.P. Belavkin, J. Phys. A {\bf 24}, 1495 (1991);
J. Dalibard, Y. Castin, and K. Molmer, Phys. Rev. Lett. {\bf 68}, 580 (1992); 
C.W. Gardiner, A.S. Perkins, and P. Zoller, Phys. Rev. A {\bf 46}, 4363 (1992); 
N. Gisin and I.C. Percival, J. Phys. A {\bf 25}, 5677 (1992); 
H. Carmichael, {\it An Open Systems Approach to Quantum Optics} (Springer, Berlin, 1993); 
R. Schack, T.A. Brun, and I.C. Percival,  J. Phys. A {\bf 28}, 5401 (1995);
H.J. Carmichael, ed., Quant. Semiclass. Opt. {\bf 8}, no. 1 (1996);
I.C. Percival, {\it Quantum State Diffusion} (Cambridge University Press, Cambridge, 1998).

\bibitem{habib} T. Bhattacharya, S. Habib, and K. Jacobs, Phys. Rev. Lett. {\bf 85}, 4852 (2000). 

\bibitem{wiseman2} H.M. Wiseman and G.J. Milburn, Phys. Rev. A {\bf 47}, 1652 (1993).

\bibitem{wiseman3} H.M. Wiseman and L. Di\'{o}si, Chem. Phys. {\bf 268}, 91 (2001).
 
\bibitem{goetsch1} P. Goetsch, P. Tombesi, and F. Haake, Phys. Rev. A {\bf 51}, 136 (1995).

\bibitem{goetsch2} P. Goetsch, P. Tombesi, and D. Vitali, Phys. Rev. A {\bf 60}, 4519 (1996).

\bibitem{giovannetti} V. Giovannetti, P. Tombesi, and D. Vitali, Phys. Rev. A {\bf 60}, 1549 (1999). 

\bibitem{halo} J.R. Anglin and W.H. Zurek, Phys. Rev. D {\bf 53}, 7327 (1996).
 
\bibitem{diosi} L. Di\'{o}si,  Europhys. Lett. {\bf 22}, 1 (1993).



\end{thebibliography}
\end{document}